\documentstyle[prl,aps, graphicx, epsf]{revtex}

\def\be{\begin{equation}}
\def\ee{\end{equation}}
\def\bea{\begin{eqnarray}}
\def\eea{\end{eqnarray}}
\begin{document}
\draft

\title{Measurements of Deuteron Photodisintegration up to 4.0 GeV}
\author{C. Bochna$^{1}$,
 B.P.~Terburg$^{1}$,
 D.J.~Abbott$^{2}$, A.~Ahmidouch$^{3}$, 
C.S.~Armstrong$^{4}$, J.~Arrington$^{5}$, 
K.A.~Assamagan$^{6}$, O.K.~Baker$^{2,6}$, S.P.~Barrow$^{7}$, 
D.P.~Beatty$^{7}$, D.H.~Beck$^{1}$, S.Y.~Beedoe$^{8}$,
E.J.~Beise$^{9}$, J.E.~Belz$^{10}$,  
P.E.~Bosted$^{11}$, 
E.J.~Brash$^{12,18}$, H.~Breuer$^{9}$, R.V.~Cadman$^{1}$,
L.~Cardman$^{2}$, R.D.~Carlini$^{2}$, J.~Cha$^{6}$, N.S.~Chant$^{9}$, 
G.~Collins$^{9}$, C.~Cothran$^{13}$, W.J.~Cummings$^{14}$,
S.~Danagoulian$^{8}$, F.A.~Duncan$^{9}$, J.A.~Dunne$^{2}$, 
D.~Dutta$^{15}$, T.~Eden$^{6}$, R.~Ent$^{2}$,
B.W.~Filippone$^{5}$,
T.A.~Forest$^{1}$, H.T.~Fortune$^{7}$, V.V.~Frolov$^{16}$, 
H.~Gao$^{1,17}$, 
D.F.~Geesaman$^{14}$, R.~Gilman$^{18}$,
P.L.J.~Gueye$^{6}$, K.K.~Gustafsson$^{9}$, J-O.~Hansen$^{14}$, 
M.~Harvey$^{6}$, W.~Hinton$^{6}$, 
R.J.~Holt$^{1}$, H.E.~Jackson$^{14}$,
C.E.~Keppel$^{6}$, M.A.~Khandaker$^{19}$, E.R.~Kinney$^{20}$, 
A.~Klein$^{21}$,
D.M.~Koltenuk$^{7}$, G. Kumbartzki$^{17}$, 
A.F.~Lung$^{9}$, D.J.~Mack$^{2}$,
R.~Madey$^{3,6},$ P.~Markowitz$^{22}$, 
K.W.~McFarlane$^{19}$, R.D.~McKeown$^{5}$, 
D.G.~Meekins$^{4}$,
Z-E.~Meziani$^{23}$, M.A.~Miller$^{1}$,
J.H.~Mitchell$^{2}$, H.G.~Mkrtchyan$^{24}$, 
R.M.~Mohring$^{9}$, 
J.~Napolitano$^{16}$,
A.M.~Nathan$^{1}$, G.~Niculescu$^{6}$, I.~Niculescu$^{6}$,
T.G.~O'Neill$^{14}$, B.R.~Owen$^{1}$, S.F.~Pate$^{25}$, 
D.H.~Potterveld$^{14}$, J.W.~Price$^{16}$, G.L.~Rakness$^{20}$,
R.~Ransome$^{18}$, 
J.~Reinhold$^{14}$, P.M.~Rutt$^{18}$,
C.~W.~Salgado$^{19}$, G.~Savage$^{6}$,
R.E.~Segel$^{15}$, N.~Simicevic$^{1}$, P.~Stoler$^{16}$, 
R.~Suleiman$^{3}$, L.~Tang$^{6}$, 
 D.~van Westrum$^{20}$,
W.F.~Vulcan$^{2}$, S.~Williamson$^{1}$,
M.T.~Witkowski$^{16}$, S.A.~Wood$^{2}$, C.~Yan$^{2}$, 
B.~Zeidman$^{14}$}

\address{$^{1}$University of Illinois at Urbana-Champaign, Urbana, Illinois 61801.
$^{2}$Thomas Jefferson National Accelerator Facility, Newport News,
Virginia 23606.
$^{3}$Kent State University, Kent, Ohio 44242.
$^{4}$College of William and Mary, Williamsburg, Virginia 23187.
$^{5}$California Institute of Technology, Pasadena, California 91125.
$^{6}$Hampton University, Hampton, Virginia 23668.
$^{7}$University of Pennsylvania, Philadelphia, Pennsylvania 19104.
$^{8}$North Carolina A\&T State University, Greensboro, North
Carolina 27411.
$^{9}$University of Maryland, College Park, Maryland, 20742.
$^{10}$TRIUMF, Vancouver, British Columbia, Canada V6T 2A3.
$^{11}$American University, Washington, D.C.~20016.
$^{12}$University of Regina, Regina, Saskatchewan, Canada S4S 0A2.
$^{13}$University of Virginia, Charlottesville, Virginia 22901.
$^{14}$Argonne National Laboratory, Argonne, Illinois 60439.
$^{15}$Northwestern University, Evanston, Illinois 60201.
$^{16}$Rensselaer Polytechnic Institute, Troy, New York 12180.
$^{17}$Massachusetts Institute of Technology, Cambridge,
Massachusetts 02139.
$^{18}$Rutgers University, New Brunswick, New Jersey 08903.
$^{19}$Norfolk State University, Norfolk, Virginia 23504.
$^{20}$University of Colorado, Boulder, Colorado 80309.
$^{21}$Old Dominion University, Norfolk, Virginia 23529.
$^{22}$Florida International University, University Park, Florida 33199.
$^{23}$Temple University, Philadelphia, Pennsylvania 19122.
$^{24}$Yerevan Physics Institute, Yerevan, Armenia.
$^{25}$New Mexico State University, Las Cruces, New Mexico 88003.
}
\date{\today}
\maketitle

\begin{abstract}
The first measurements of the differential cross section
for the $d(\gamma,p)n$ reaction up to 4.0 GeV were performed 
at Continuous Electron Beam Accelerator Facility (CEBAF) at Jefferson Lab. 
We report the cross sections at the proton center-of-mass angles of
36$^\circ$, 52$^\circ$, 69$^\circ$ and 89$^\circ$. These results are
in reasonable agreement with previous measurements at lower energy.  
The $89^\circ$ and $69^\circ$ data show constituent-counting-rule
behavior up to 4.0 GeV photon energy. The $36^\circ$ and $52^\circ$
data disagree with the counting rule behavior.  The quantum
chromodynamics (QCD) model of
nuclear reactions involving reduced amplitudes disagrees with the
present data.
\end{abstract}

\narrowtext
\twocolumn

To reconcile low energy and high energy descriptions of hadronic
matter, nuclear physics must determine when it is justified to make a
transition from meson-nucleon degrees of freedom to quark-gluon
degrees of freedom in the description of a nuclear reaction. The QCD
content of nuclei was studied first by Brodsky and Chertok
\cite{chertok}. A possible signature for this transition is that the
reaction cross section begins to scale\footnote{Scaling in this
  context implies a dependence on a reduced set of kinematic variables
  indicating a simplification of the reaction dynamics.} at some
incident energy.  If scaling were indeed observed, characterization of
the approach to scaling would be essential to understand how the
dynamics are simplified. High energy two-body photodisintegration of
the deuteron ($\gamma d \rightarrow p n$) is particularly well suited
for these studies because it is amenable to theoretical calculation
and relatively high momentum transfer to the constituents can be
achieved at relatively modest photon energies \cite{holt}.

Previous measurements \cite{slac,belz} 
for the $d(\gamma,p)n$ reaction indicate the onset
of scaling behavior at a photon energy of 1 GeV at a reaction angle of
$\theta_{cm} =90^\circ$; however, this limited data set does not show
scaling at other reaction angles. Measurement of the angular
distributions of scaling thresholds is an important part of
characterizing the reaction process.  In this Letter, we present new
data to address this issue. 
                                      
Predictions for the energy dependence of the cross section at high
energies are given by constituent counting rules \cite{counting}, the
reduced nuclear amplitude analysis (RNA) \cite{rna}, the asymptotic
meson exchange model (AMEC) \cite{nagornyi} and the quark-gluon string
model (QGS) \cite{string}. Constituent counting rules predict that the
energy dependence for the two-body exclusive reaction cross section
should be given by  
\begin{equation} 
d\sigma/dt = \frac{h(\theta_{cm})}{s^{n-2}},
\end{equation}
where the Mandelstam variables $s$ and $t$ are the square of the total
energy in the center-of-mass frame and the momentum transfer squared
in the $s$ channel respectively. The symbol $n$ denotes the total
number of elementary fields in the initial and final states, and $n-2$
is 11 for the $d(\gamma,p)n$ reaction. The quantity $h(\theta_{cm})$
depends on details of the dynamics of the process. Constituent
counting rules are believed valid at energies much greater than the
masses of the participating particles, perhaps in the perturbative QCD
region. The previous data at $90^\circ$ scale according to these rules 
with a photon energy threshold of only 1 GeV.  

In the RNA approach, the amplitude is described in terms of parton
exchange between the two nucleons. The low energy components
responsible for quark binding within the nucleons are removed by
dividing out the empirical nucleon form factors.  While the reduced
nuclear amplitude analysis appears to describe the electron-deuteron
elastic scattering cross section above a momentum transfer of 1
GeV/c\cite{chertok}, it does not give a good description of the
previous deuteron photodisintegration data. This is surprising because
this model is expected to approach scaling at lower energies than the
constituent counting rules. The QGS model is based on Regge
phenomenology and is expected to be valid at small values of $t$ where
the parameters are best determined. Thus, the QGS model is expected to
be valid at small reaction angles, while most of the existing data is
at large angles. 

The traditional meson exchange models \cite{lee,laget} describe the
data at energies below 1 GeV, but are problematic above 1 GeV.
However, the AMEC \cite{nagornyi} departs from the conventional
approach in that an asymptotic description of the nucleon-nucleon
interaction is used \cite{gross}. Although this model appears to
predict the observed energy dependence at $\theta_{cm} =90^\circ$, 
it cannot yet reproduce the magnitude of the cross section.

In this Letter we present new results from the $d(\gamma,p)n$ reaction
at proton center-of-mass angles of $36^\circ$, $52^\circ$, $69^\circ$
and $89^\circ$. These data overlap with existing SLAC measurements,
extending them to higher energy and providing more complete angular
coverage. 

Because the cross sections are much less than 1 nb/sr, it was
essential to have a high-current, high duty factor electron beam of
multi-GeV energy, and to use well-shielded spectrometers of large solid
angle. This was achieved by performing the experiment in Hall C at
Thomas Jefferson National Accelerator Facility (Jefferson Lab). A 25
$\mu$A continuous (CW) electron beam in the energy range 0.845 to
4.045 GeV in steps of 0.8 GeV was incident on a $6\%$ copper radiator
to create an untagged photon beam. The resulting electron and
bremsstrahlung beam impinged on a 15-cm liquid deuterium
target. Because the $\gamma d \rightarrow p n$ reaction is a two-body
process, the photon energy could be reconstructed from the measured
final-state proton momentum and scattering angle. Events with pion
production were excluded by accepting only the protons with the
highest momenta. Protons were detected with the High Momentum
Spectrometer (HMS) \cite{hms}.  

The HMS had a solid angle of 6.8 msr and a momentum acceptance of
$18\%$. The detector system consisted of a plastic scintillator
hodoscope, drift chambers, and a gas \v{C}erenkov counter.  The
hodoscope was used to form a trigger and also to provide the
time-of-flight information for particle identification.  Drift
chambers were employed to measure particle trajectories in order to
calculate the momenta and reaction angles. It is important to
distinguish protons from pions and deuterons. A gas \v{C}erenkov
counter, filled with $C_4F_{10}$ gas, was used in combination with
time-of-flight (TOF) information to separate pions and protons above
2.8 GeV/c, while TOF alone was used for pion rejection at lower
momenta. Deuteron rejection was accomplished by a TOF cut at all
momenta. 
   
Background contributions from the target windows were removed by
placing cuts on the reconstructed target position and subtracting the
yield obtained with a cell of identical dimensions that was either
empty or filled with liquid hydrogen to simulate bremsstrahlung in the
deuterium. This target length cut led to a $3\%$ error for the target
thickness. Deuterium and empty target data were taken alternately
during the experiment. Data were taken with the hydrogen target at
some of the kinematic settings to cross check the procedure of the
empty target subtraction. The yield from electrodisintegration was
measured by repeating the procedure without the radiator present. This
yield was treated as a background and was subtracted from the
photodisintegration yield with an energy-dependent correction factor
to take into account the modification of the electron beam flux and
energy distribution by the radiator~\cite{belz2}. An additional
systematic error of $3\%$ was assigned to this background subtraction
procedure.

The photon energy bin limits were chosen to kinematically eliminate
protons from photopion production processes, and to eliminate the
bremsstrahlung end point, for which the photon flux is less well
known. This photon energy bin varied in width from 60 MeV at
$36^\circ$ to 115 MeV at $89^\circ$. The bremsstrahlung photon flux
was calculated with an estimated $3\%$ uncertainty using the
thick-target bremsstrahlung computer code developed by
Belz~\cite{belz2} , which was cross-checked by using the codes of
Matthews and Owens~\cite{thick}. The spectrometer solid angle for the
extended target was studied by comparing a Monte Carlo simulation with
measurements made with both a movable solid carbon target and the
extended cryotargets.  In addition, the dependence of the results on
the choice of optics parameters for the HMS was studied. An overall
error of $7\%$ in the HMS acceptance was determined from these
studies. A proton absorption correction was applied to compensate for
the scattering in the spectrometer windows and the detector
stack. This proton attenuation was measured \cite{vanWestrum} to be
$(5.5 \pm 2.0)\  \%$ by comparing singles $H(e,e')$ and coincidence
$H(e,e'p)$ measurements. Corrections were also applied for the
computer deadtime and the tracking efficiency.

The overall systematic uncertainty is found to be $\le 11.5\%$. The
uncertainties from the beam current measurement, beam energy
determination and photon energy reconstruction in the measured
quantity $s^{11}d\sigma/dt$ were less than $3\%$. The uncertainty from
the particle identification is $\le 5\%$. An additional systematic
error of $3\%$ resulted from using a reduced solid angle in the HMS to
avoid an obstruction from an HMS vacuum valve that was inadvertantly
partially closed\cite{bochna,terburg}. Thus, only half of the HMS
solid angle was used in the analysis of the data. A separate
experiment without the obstruction verified that there was good
agreement between the half acceptance of the experiment with the
obstruction and the corresponding half acceptance for the experiment
without the obstruction. A Monte Carlo simulation was used to
determine the correction for ``in-scattered" protons from the valve
into the open half of the spectrometer solid angle. The amount of this
correction is less than $2\%$ in the worst case.  This selection of
half the spectrometer solid angle resulted in a reduction of the
observed reaction center of mass angle by approximately $1^\circ$ from
the central spectrometer setting. 

Figure 1 shows the world data of $s^{11}d\sigma/dt$ above $\sim$ 0.4
GeV for the $\gamma d \rightarrow p n$ reaction at 
$\theta_{cm}= 36^\circ, 52^\circ, 69^\circ$ and $89^\circ$ as a
function of the photon energy. The $89^\circ$ data are shown in the
top panel of Fig.~1. The SLAC NE17 data \cite{belz} exhibit 
scaling behavior starting at photon energies around 1 GeV, 
corresponding to proton transverse momenta ($p_{T}$) of 
$\sim 1.0 \ {\rm GeV/c}$. For the $d(\gamma,p)n$ 
reaction, ${p_{T}}^{2}$ can be expressed as
${\frac{1}{2}}E_{\gamma}M_{d}{sin^{2}(\theta_{cm})}$ with $M_{d}$
being the deuteron mass. The Jefferson Lab data are in good agreement
with the previous SLAC data. The new data continue to show the scaling
behavior up to the highest photon energy of 4.0 GeV. The differential
cross section $d\sigma/dt$ from the present work at photon energies
above 1.0 GeV were fitted with the form of $A/{s^{n-2}}$, and 
$n-2 = 11.1 \pm 0.3$ was obtained from the fit. It is surprising that
the counting rule appears to work so well considering the fact that
the momentum transfer to the individual quark is below 1.0 
${\rm  (GeV/c)}^2$ for these measurements, where the strong coupling
constant still varies significantly as a function of the momentum
transfer.

The solid line in Fig.~1 represents the meson-exchange calculation of
Lee~\cite{lee}, which is a standard calculation that reproduces the
measured $NN$ phase shifts up to 2.0 GeV and is also constrained by
photomeson \cite{lee2} production data. The calculation gives a
reasonable description of the data below 0.5 GeV, but deviates above
1.0 GeV. The meson-exchange model calculation of Laget \cite{laget}
(not shown) reproduces the experimental data fairly well from the
threshold up to about 0.6 GeV. The long-dashed line represents a
reduced nuclear amplitude analysis~\cite{rna} with a normalization
factor chosen to agree with the data at 
$E_{\gamma} = 1.6\ {\rm  GeV}$. This RNA curve falls below 
the high $E_{\gamma}$ data and
does not reach an asymptotic limit at these energies. The dotted line
is a calculation (AMEC) by Nagorny\u{i} {\it et al.} \cite{nagornyi}
with the normalization performed at 1.0 GeV. Although this calculation
gives a different energy dependence to the cross section than that of
the constituent counting rule, it does give an energy dependence very
close to $1/s^{11}$ in the energy region of 1.5 to 3.5 GeV. More
calculations of this type are necessary at the other angles. The QGS
calculation \cite{string} is shown as the dash-dotted line. It
deviates from the data at large angles as expected.

The data at $52^\circ$ and $69^\circ$ are shown in the center panels
of Fig.~1. The $69^\circ$ data appear to scale according to the
constituent counting rules.  A fit to this cross section gives 
$n-2 = 10.9 \pm 0.2$ in good agreement with that at $89^\circ$.  At
$52^\circ$ the present data are in reasonable agreement with the
previous SLAC measurement and have significantly improved statistical
errors. The results deviate from the scaling behavior predicted by
counting rules. A fit of the scaling formula to the cross section data
gives $n-2 = 9.5 \pm 0.2$. The  QGS calculation \cite{string} is in
reasonable agreement with the data in the photon energy region between
1.0 to 4.0 GeV. 

The bottom panel in Fig.~1 shows the data at
$\theta_{cm}=36^\circ$. Fitting the scaling formula to the Jefferson
Lab cross section data above a photon energy of 1 GeV gives 
$n-2=9.6 \pm 0.1$. The highest $p_{T}$ from NE17 measurements at this
$\theta_{cm}$ is 0.7 GeV/c. The experimental uncertainties and the
limited energy region covered by the NE17 experiment preclude any
conclusion with regard to the scaling behavior in the photon energy
range below 3 GeV. The Jefferson Lab data are in reasonable agreement
with the SLAC data~\cite{belz} when the large angular dependence at
forward angles is taken into account.  The SLAC data were taken at
$36.7^\circ$, while the present data were recorded from $35.4^\circ$
at the highest energy to $36.2^\circ$ at the lowest energy. The
previous data shown in Fig.~1 are not corrected for this difference in
angles.

Although a photon energy of 3.1 GeV at $36^\circ$ corresponds to the
same $p_{T}$ where the onset of $s^{-11}$ scaling is observed at
$90^\circ$, there is no evidence from the $36^\circ$ data that scaling
has set in (see arrows in Fig.~1). The RNA \cite{rna} deviates
significantly from the data at $36^\circ$, but the QGS calculation
\cite{string} gives a reasonable energy dependence. 

To summarize, the data at $\theta_{cm}= 89-90^\circ$ continue to show
scaling behavior up to 4 GeV.  The first evidence is observed for a
similar scaling behavior at $69^\circ$. The asymptotic meson exchange
model is in fair agreement with the $90^\circ$ data in the photon
energy region of 1.0 to 3.0 GeV. Thus far, no constituent counting
scaling behavior is observed for the $36^\circ$ or $52^\circ$
data. The new data at $36^\circ$ and $52^\circ$ rule out a scaling
threshold of $p_{T} = 1.0\ {\rm (GeV/c)}^2$. The reduced nuclear
amplitude analysis does not agree with the present data.  The QGS
model has an energy dependence that is in reasonable agreement with
the present data at $36^\circ$ and $52^\circ$. Further measurements at
higher energies at the forward angles will test whether scaling has
really set in  and whether $p_{T}$ is the physical quantity governing
the onset of the scaling behavior. Experimental determination of the
physical observable associated with the onset of scaling is essential
for understanding the underlying mechanism for the scaling behavior in
the exclusive two-body process. 

We acknowledge the outstanding work of the staff of the accelerator
division at the Thomas Jefferson National Accelerator Facility in
delivering the high quality electron beam to 4 GeV and the Hall C
technical staff for constructing the spectrometers and the
cryotarget. We thank E.~Schulte for bremsstrahlung calculations. This
work is supported in part by the research grants from the
U.S. Department of Energy and the U.S. National Science Foundation.

% figure follows here
\begin{figure}[hbtp]
  \begin{center}
    \includegraphics[angle=0,width=3.2in]{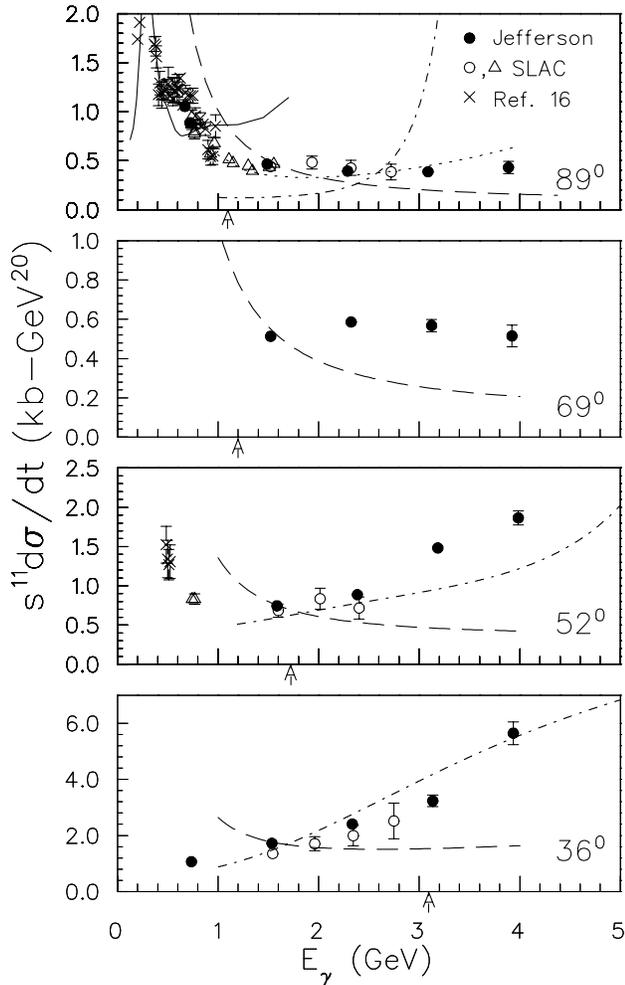}
    \caption{$s^{11}d\sigma/dt$ vs.~$E_\gamma$. 
      The present work is shown as solid circles with statistical
      uncertainties only, the SLAC NE17 data are shown as open
      circles, the SLAC NE8 data are shown as open triangles, and
      other existing data \protect\cite{other} are shown as
      crosses. The solid line is the meson-exchange model calculation
      of Lee \protect\cite{lee}. The long-dashed line is the RNA
      analysis \protect\cite{rna}, and the dotted line is
      Nagorny\u{i}'s \protect\cite{nagornyi} asymptotic meson-exchange
      calculation. The dash-dotted line is the QGS calculation
      \protect\cite{string}. The arrows indicate the photon energies
      where ${p_{T}}^{2} = 1.0\ {\rm (GeV/c)}^2$. The previous data
      are shown above at nominal center of mass angles of $37^\circ$,
      $53^\circ$, and $90^\circ$.} 
  \end{center}
\end{figure}

\end{document}